\pdfoutput=1

\documentclass[prl,twocolumn,showpacs,preprintnumbers,amsmath,amssymb,superscriptaddress,nofootinbib]{revtex4-1}
\usepackage{graphicx}

\renewcommand{\tilde}{\widetilde}
\newcommand{\mi}{\raisebox{0.75pt}{\scalebox{0.75}{$\,-\,$}}}
\renewcommand{\pl}{\raisebox{0.75pt}{\scalebox{0.75}{$\,+\,$}}}

\renewcommand{\bbox}[1]{\raisebox{-2.25pt}{\,\scalebox{1.75}{$#1$}}\,}
\newcommand{\fwbox}[2]{\text{\makebox[#1][c]{$\hspace{-150pt}\displaystyle#2\hspace{-150pt}$}}}

\newcommand{\eq}[1]{\vspace{-0.pt}\begin{equation}\hspace{-100pt}#1\hspace{-100pt}\vspace{-0.pt}\end{equation}}
\newcommand{\eqs}[1]{\vspace{-0.0pt}\begin{equation}\begin{split}#1\end{split}\vspace{-0.0pt}\end{equation}}
\newcommand{\fig}[1]{\raisebox{-26.95pt}{\includegraphics[scale=1]{#1}}}
\renewcommand{\phi}{\varphi}

\newcommand{\vecdot}[2]{#1\!\cdot\!#2}
\newcommand{\pv}[1]{\mathcal{P}#1}
\newcommand{\ie}[1]{i\epsilon_{#1}}
\newcommand{\sgn}[1]{\text{sign}(#1)}

\begin{document}

\title{New Representations of the Perturbative S-Matrix}
\author{Christian~Baadsgaard}\affiliation{Department of Physics, Princeton University, Princeton, New Jersey, USA}
\affiliation{Niels Bohr International Academy \& Discovery Center, Niels Bohr Institute, University of Copenhagen, Denmark}
\author{N.~E.~J.~Bjerrum-Bohr}\author{Jacob~L.~Bourjaily}\author{Simon~Caron-Huot}\author{Poul~H.~Damgaard}\affiliation{Niels Bohr International Academy \& Discovery Center, Niels Bohr Institute, University of Copenhagen, Denmark}
\author{Bo~Feng}\affiliation{Zhejiang Institute of Modern Physics, Zhejiang University, Hangzhou, PR China}
\date{\today}

\begin{abstract}
We propose a new framework to represent the perturbative S-matrix which is well-defined for all quantum field theories of massless particles, constructed from tree-level amplitudes and integrable term-by-term. This representation is derived from the Feynman expansion through a series of partial fraction identities, discarding terms that vanish upon integration. Loop integrands are expressed in terms of ``\mbox{$Q$-cuts}'' that involve both off-shell and on-shell loop-momenta, defined with a precise contour prescription that can be evaluated by ordinary methods. This framework implies recent results found in the scattering equation formalism at one-loop, and it has a natural extension to all orders---even non-planar theories without well-defined forward limits or good ultraviolet behavior.
\end{abstract}
\maketitle

\section{Introduction}\label{introduction_section}\vspace{-10pt}

Since the revolutionary developments in quantum field theory around the middle of the last century, Feynman diagrams have been an essential tool for computing scattering amplitudes. As represented by the Feynman expansion, amplitudes can be determined perturbatively by summing over all graphs with a fixed number of loops that connect the external states by appropriate vertices and propagators and integrating over all the internal loop-momenta. While intuitive and ultimately correct, the Feynman expansion rapidly becomes intractable both because the number of diagrams grows rapidly with the number of external legs and loops, but also because individual diagrams in many theories depend on a great deal of purely theoretical data (such as gauge redundancies) that have no consequences for physical predictions.

Because of this, there has long been interest in finding formulations of perturbation theory without any explicit reference to unphysical data. This was the principal motivation behind the Feynman tree theorem~\cite{Feynman:1963ax}, the essence of which can be be understood (in a rather novel way) from the well-known partial fraction identity:
\vspace{-5pt}\eq{\frac{1}{D_1\!\cdots D_m}=\sum_{i=1}^m\frac{1}{D_i}\Bigg[\prod_{j\neq i}\frac{1}{D_{j}-D_{i}}\Bigg].\vspace{-5pt}\label{partial_fraction_identity}}
When the factors $1/D_i$ are Feynman propagators for an off-shell loop momentum $\ell$, the terms $(D_j\mi D_i)$ become linear in $\ell$ and can therefore be interpreted as propagators involving an {\it on-shell} momentum $\ell$. Expanding every term in the Feynman expansion in this way, it is tempting to view the coefficient of each $1/D_i(\ell)$ as a lower-loop {\it amplitude}, evaluated in the forward-limit. This is precisely correct for theories (such as those with supersymmetry) whose amplitudes are finite in the forward-limit; and this leads, for example, to the representation found in \mbox{ref.\ \cite{Geyer:2015bja}} for one-loop amplitudes in supergravity, derived within the scattering equation formalism \cite{Cachazo:2013hca}. But for most theories, however, scattering amplitudes diverge in the forward-limit, preventing any (unqualified) interpretation of the terms in this expansion as limits of lower-loop {amplitudes}. 

In this Letter, we present a new representation of amplitudes which avoids this obstruction, allowing one to write the coefficient of each off-shell propagator $1/D_i(\ell)$ in the expansion above in terms of ``\mbox{$Q$-cuts}'', which are well-defined in any theory in terms of gauge-invariant tree-amplitudes alone. In the following section, we derive this representation in detail for one-loop amplitudes and show how each term in this expansion can be integrated by ordinary means. We then describe how this approach generalizes to higher loops, illustrating the rich structure that emerges beyond two-loops.

\vspace{-10pt}\section{The $Q$-Cut Representation At One-Loop}\vspace{-10pt}\label{one_loop_section}

For the sake of concreteness and clarity, let us restrict our attention to one-loop amplitudes of theories with massless particles, so that Feynman propagators for the loop momenta are of the form $N(\ell)/(\ell\pl P)^2$. Both in order to generalize the partial fraction decomposition to cases with loop-dependent numerators, and to introduce ideas that will prove useful later on, let us derive equation (\ref{partial_fraction_identity}) as an instance of Cauchy's residue theorem. Consider transforming $\ell\!\mapsto\!\ell\pl\eta$ where $\eta$ is orthogonal to the external momenta\footnote{This is always possible in the framework of dimensional regularization by taking $\eta$ to lie along the extra dimensions.} satisfying $\eta^2\!=\!z$ so that $\ell^2\!\mapsto\!\ell^2\pl z$, and $D_i\!\mapsto\!D_i\pl z$ for propagators involving $\ell$. Transforming a Feynman diagram this way and dividing by $z$ results in
\vspace{-5pt}\eq{\frac{N(\ell)}{D_1(\ell)\!\cdots D_m(\ell)}\mapsto\frac{1}{z}\frac{N(\ell,z)}{(D_1(\ell)\!+z)\cdots(D_m(\ell)\!+z)}.\vspace{-5pt}\label{cauchy_theorem_precursor_to_partial_fractions}}
After this deformation, the statement that the sum of all the residues in $z$ vanishes trivially reduces to equation (\ref{partial_fraction_identity}) as a special case.

Using this generalization of the partial fraction expansion to decompose every integral of the Feynman expansion, it is easy to see that every term has the form (up to a shift in $\ell$ by external momenta),
\vspace{-5pt}\eq{\frac{1}{\ell^2}\left[\frac{N(\ell)}{(2\vecdot{\ell}{P_1}\pl Q_1)\cdots(2\vecdot{\ell}{P_m}\pl Q_m)}\right]
,\vspace{-5pt}\label{form_of_linearized_ints}}
where $N(\ell)$ accounts for both the numerators of the diagrams and any loop-independent propagators. Let $\mathcal{I}(\ell)$ denote the factors in the square brackets above.

Although forward-limit divergences prevent us from interpreting $\mathcal{I}(\ell)$ in terms of an entire tree-amplitude in general, it turns out that we can construct $\mathcal{I}(\ell)$ in terms of tree-level objects {\it up to terms that vanish upon integration}. This becomes possible after one further partial-fraction-like expansion---this time, in the scale of $\ell$. Concretely, consider the residue theorem resulting from:
\vspace{-5pt}\eq{\mathcal{I}(\ell)\!\mapsto\!\widetilde{\mathcal{I}}(\alpha,\ell)\!\equiv\!\frac{\widetilde{\mathcal{I}}(\alpha\ell)}{(\alpha\mi1)}\,.\vspace{-5pt}}
Clearly, $\mathcal{I}(\ell)$ is recovered as the residue of $\widetilde{\mathcal{I}}(\alpha,\ell)$ at $\alpha\!=\!1$; and by Cauchy's theorem, this is equal to (minus) the sum of all other residues. These residues are associated with three types of poles: at zero, at infinity, and at finite locations ($\alpha\!\neq\!1$). By inspection of equation (\ref{form_of_linearized_ints}), residues at $\alpha\!=\!0$ correspond to integrals of the form,
\vspace{-2.5pt}\eq{\int\!\!d^d\ell\,\,\frac{1}{\ell^2}\widetilde{N}(\ell)\prod_k\frac{1}{2\vecdot{\ell}{P_k}}
~\bbox{\Rightarrow}~0\,,\vspace{-5pt}}
where the product runs over only those factors for which $Q_k\!=\!0$, and $\widetilde{N}(\ell)$ denotes all other factors at $\alpha\!=\!0$. Integrals of this form must vanish upon integration in any number of dimensions $d$ because the denominator is homogeneous in $\ell$ and hence scale-free. Similarly, the Laurent expansion of $\widetilde{\mathcal{I}}(\alpha,\ell)$ at $\alpha\!\to\!\infty$ can involve only terms homogeneous in $\ell$, which hence vanish upon integration. Notice that these residues precisely correspond to the terms poorly defined in the forward-limit. 

Therefore, we can replace $\mathcal{I}(\ell)$ by the sum of residues of $\widetilde{\mathcal{I}}(\alpha,\ell)$ at \mbox{$\alpha\!\notin\!\{0,1,\infty\}$.} Importantly, all such residues can be interpreted as involving {\it two} additional on-shell particles, with specific momenta determined by the successive residues. It is not hard to see that expanding every term in the Feynman expansion in this way, the coefficient of each pair of propagators becomes a product of complete tree-amplitudes evaluated for particular on-shell, internal momenta (and summing over states):
\vspace{-5pt}\eq{\hspace{-10pt}\mathcal{A}_L(\cdots,\tilde{\ell}_L,\mi\tilde{\ell}_R)\frac{1}{\ell^2}\frac{1}{\big(2\vecdot{\ell}{P_L}+P_L^2\big)}\mathcal{A}_R(\tilde{\ell}_R,\mi\tilde{\ell}_L,\cdots),\vspace{-5pt}\label{one_loop_q_cut}}
with $\tilde{\ell}_L\!\equiv\!\alpha(\ell+\eta)$ and $\tilde{\ell}_R\!\equiv\!\tilde{\ell}_L\!\pl P_L$, with $\eta^2=-\ell^2$ and
\mbox{$\alpha\!\equiv\!\mi P_L^2/(2\vecdot{\ell}{P_L})\!\neq\!0$}, and where $P_L$ denotes the sum of momenta over a partition of external legs. We refer to functions of the form of (\ref{one_loop_q_cut}) as \mbox{$Q$-cuts}, which we can represent graphically as follows:
\vspace{-7.5pt}\eq{\fig{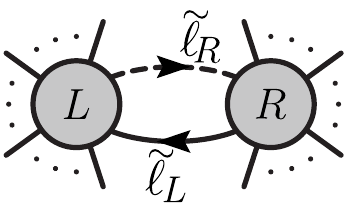}\vspace{-7.5pt}\label{one_loop_q_cut_figure}}
Notice that the shifted propagator, corresponding to the factor $1/((\ell\pl P_L)^2\mi \ell^2)$ in equation (\ref{one_loop_q_cut}), is indicated by a dashed line in the figure above to distinguish it from the unshifted, off-shell propagator $1/\ell^2$.

We claim that the sum over all \mbox{$Q$-cuts} (with \mbox{$P_L^2\!\neq\!0$}) reproduces any one-loop amplitude. Notice that the integrand of a \mbox{$Q$-cut} is similar to a Cutkosky unitarity cut~\cite{Cutkosky:1960sp}. The principal novelty involved in the \mbox{$Q$-cut} is that the amplitudes involved are evaluated with {\it shifted} (on-shell) values of $(\tilde{\ell}_L,\tilde{\ell}_R)$, multiplied by unusual propagators. 

\vspace{-10pt}\section{Contours of integration}\vspace{-10pt}\label{loop_contour_section}
At a fundamental level, the causal structure of scattering amplitudes is encoded in the Feynman $i\epsilon$-prescription, critical to the precise definition of the loop integration contour. It will be useful here to observe that every Feynman propagator can be assigned its own $\epsilon$, transforming $D_j\!\mapsto\!D_j\pl\ie{j}$; so long as each $\epsilon_j$ is real and positive, the physical contour will be unambiguous (and independent of the $\epsilon_j$'s). 

Since \mbox{$Q$-cuts} do not involve products of Feynman propagators, it is not immediately clear how to assign $i\epsilon$'s to the linear poles appearing in their definition. However, if we had started with a single Feynman integral with specific $\epsilon$'s for each propagator, then the partial fraction expansion would result in terms of the form $D_{ij}\!\equiv\!(D_i\mi D_j)$, with contours prescribed by shifts involving $\ie{ij}\!\equiv\!i(\epsilon_i\mi\epsilon_j)$, the signs of which will be fully determined by the (arbitrary) ordering of the original $\epsilon$'s. This always provides a precise contour of integration for the resulting expressions that is guaranteed to match the original expression. The problem is in going in the other direction: to assign an unambiguous prescription for the $\epsilon$'s associated with the linear-factors in $\ell$ of each \mbox{$Q$-cut} integral.\footnote{In planar theories, where all the $\epsilon$'s of all propagators can be identified and conventionally ordered according to the ordering of the external legs, an unambiguous convention for the signs of the $\epsilon$'s for any linear factor can be easily assigned.}

(We should mention that the contour we describe here requires that on-shell tree amplitudes are represented in a way that involves only local poles. Representations of trees generated by the BCFW recursion relations~\cite{Britto:2005fq}, for example, involve spurious, complex poles in individual terms. Finding a contour prescription for such terms is an important and interesting open problem.)

It turns out that a maximally democratic contour prescription will always work---meaning, that it is guaranteed to match the Feynman contour. Specifically, we may simply average over the possible relative orderings of the $\epsilon_j$'s for the initial propagators.

It will help to clarify this contour prescription with a few examples. Consider a single propagator linear in $\ell$, denoted $1/x$. Its $\epsilon$ can be arbitrary, so we are instructed average over the $2!$ possible choices for the sign of its $i\epsilon$-prescription; thus, the contour prescription becomes the principle value via the replacement:
\vspace{-5pt}\eq{\frac{1}{x}\mapsto\frac{1}{2}\left(\frac{1}{x\pl i\epsilon}\pl\frac{1}{x\mi i\epsilon}\right)\equiv\pv{\Big[\frac{1}{x}\Big]}.\vspace{-5pt}\label{single_factor_pv}}
(Recall the possible contour prescriptions---defined by,
\vspace{-5pt}\eq{\frac{1}{x\pm i\epsilon}\equiv\pv{\Big[\frac{1}{x}\Big]}\mp i \pi \delta(x),\vspace{-5pt}}
where $\delta(x)$ is the (non-holomorphic) Dirac $\delta$-function.)

For three propagators (the case involving two linear propagators after the partial fraction decomposition), the result is rather less trivial: the $3!$ relative orderings for the $\epsilon$'s of the three propagators result in a more interesting distribution of differences. It is a simple exercise to see that the resulting contour prescription for the any pair of differences should be:
\vspace{-5pt}\eq{\frac{1}{xy}\mapsto\pv{\Big[\frac{1}{x}\Big]}\pv{\Big[\frac{1}{y}\Big]}-\frac{1}{3}\pi^2\delta(x)\delta(y).\vspace{-5pt}}
For four propagators (three linear ones), the prescription is quite similar: $1/(xyz)$ should be replaced by,
\vspace{-20pt}\eqs{\hspace{-60pt}&\\\hspace{-50pt}&\phantom{-\,}\pv{\Big[\frac{1}{x}\Big]}\pv{\Big[\frac{1}{y}\Big]}\pv{\Big[\frac{1}{z}\Big]}\\[0pt]\hspace{-20pt}&-\frac{1}{3}\pi^2\Big(\delta(x)\delta(y)\pv{\Big[\frac{1}{z}\Big]}\pl\delta(y)\delta(z)\pv{\Big[\frac{1}{x}\Big]}\pl\delta(z)\delta(x)\pv{\Big[\frac{1}{y}\Big]}\Big).\hspace{-30pt}\\[-20pt]~\hspace{-20pt}\label{three_factor_pv}}
And for five or more, there will be terms involving four $\delta$-functions for each quadruplet (times $\pl\pi^4/5$), etc.

\section{One-Loop Examples}\label{one_loop_examples_subsection}

To exemplify the contour prescription, consider a simple (massive) bubble integral with two propagators. After partial fractioning, and using the contour prescription described above, it becomes:
\vspace{-5pt}\eq{\int\!\!\!\frac{d^d\ell}{\pi^{d/2}}\,\,\frac{1}{\ell^2+i\epsilon} \pv{\Big[\frac{2}{2\ell{\cdot}p+p^2}\Big]}\,.~~~~\label{bubble}\vspace{-5pt}}
To simplify the $\ell$-integration, it is possible to combine the denominators via Schwinger parameters, as usual. The following identities, which generalize the usual distribution formula, $i/(x\pl i\epsilon)\!\Leftrightarrow\!\int_0^\infty\!\!da\, e^{i\;\!a\;\!x}$, are useful:
\vspace{-5pt}\eq{\pv{\Big[\frac{2i}{x}\Big]}\Leftrightarrow\int_{-\infty}^{\infty} \!\!\!\!\!da\, \,\sgn{a}\,e^{i \;\!a\;\!x},~
2\pi\delta(x)\Leftrightarrow\!\int_{-\infty}^{\infty} \!\!\!\!\!da\, \,e^{i\;\!a\;\!x}.~~~\label{schwinger}~~\vspace{-5pt}
}
Integrating over $\ell$ and an overall Schwinger parameter,  the integral (\ref{bubble}) readily becomes:
\vspace{-5pt}\eq{
\Gamma\Big(2-\frac{d}{2}\Big)\int_{-\infty}^\infty \!\!\!\!\!da\,\, \frac{\sgn{a}}{(-a(1-a)p^2-i\epsilon)^{2-d/2}}\,.
\label{bubble_with_feynman_parameter}~~~~~~~~~~\vspace{-5pt}}
One can see that the regions $a\!>\!1$ and $a\!<\!0$ exactly cancel each other,
leaving the region $0\!<\!a\!<\!1$, which precisely reproduces the usual Feynman-parameter expression for this integral.
We could similarly prove, by exploiting cancelations in Schwinger parameter space, that any Feynman integral is correctly reproduced by the sum of its \mbox{$Q$-cuts}, integrated
using the contour prescribed above.

As a relatively simple but illustrative application, consider the amplitude for incoming gluons with the same ($\!\pl\!$)-helicity in planar Yang-Mills theory. 
Considering for simplicity the contribution from a complex scalar loop, and substituting the product of trees using the notation of \mbox{ref.~\cite{Bern:1995db}}, the \mbox{$Q$-cut} defined in equation (\ref{one_loop_q_cut}) becomes:
\vspace{-5pt}\eq{
 \left(\!\frac{[12]}{\langle 12\rangle}\frac{\mu^2{-}\ell^2}{(2\vecdot{\ell}{p_1})}\!\right)\!
\frac{1}{\ell^2}\frac{\big[p_{12}^2/(2\vecdot{\ell}{p_{12}})\big]^2}{(2\vecdot{\ell}{p_{12}}\pl p_{12}^2)}\!
\left(\!\frac{[34]}{\langle34\rangle} \frac{\mu^2{-}\ell^2}{(\!\mi2\vecdot{\ell}{p_4})}\!\right).~~~~~~
 \label{integrand_all_plus}\vspace{-5pt}
}
(The full amplitude will be represented by the sum of this \mbox{$Q$-cut} and its three cyclic rotations.) For this helicity amplitude, an identical result would be obtained for a gluon in the
loop.\footnote{So long as only physical ($d$-dimensional) gluons are included in the polarization sums, then no ghosts are needed.} 
Up to integrals that vanish upon integration, the preceding is equivalent to,
\vspace{-5pt}\eq{
\frac{[12][34]}{\langle 12\rangle\langle34\rangle}
\frac{\mu^4}{\ell^2(2\ell{\cdot}p_1)(2\ell{\cdot}p_{12}{+}p_{12}^2)(-2\ell{\cdot}p_4)}\,. \vspace{-5pt}\label{integrand_all_plus_simplified}
}
In the conventional unitarity method~\cite{Bern:1994zx,Bern:1994cg}, the two-particle cut of this is recognized as that of a box integral with $\mu^4$ numerator, which is then integrated. 
Similarly, its single-cut matches that appearing in the Feynman tree theorem (see \mbox{e.g.\ \cite{Catani:2008xa,NigelGlover:2008ur}}).
The distinctive feature of the present approach, however, is that each term can be integrated directly, bypassing the reconstruction of the off-shell integrand, and without requiring a forward-limit interpretation of the single-cut.

Specifically, paying attention to the contour in equation (\ref{three_factor_pv}) and applying Schwinger parameters~(\ref{schwinger}) to the linear propagators, the integral over the \mbox{$Q$-cut} gives:
\vspace{-5pt}\eqs{~&
\frac{\mi i\phantom{\mi}}{16\pi^2}\frac{[12][34]}{\langle 12\rangle\langle34\rangle}
\int_{-\infty}^\infty \frac{da\,db\,dc\,\big[1+\mathcal{O}(d\mi4)\big]}{(b(1\mi a\mi b\mi c)s\pl a\,c\,t-i\epsilon)^{2-d/2}}
\nonumber\\ &\hskip-0.2cm\times \!\frac18\!\Big[ \sgn{a}\sgn{b}\sgn{c}\!+\!\frac13(\sgn{a}\pl\sgn{b}\pl\sgn{c})\Big],\\[-20pt]~
}
where $s\!\equiv\!p_{12}^2$ and $t\!\equiv\!p_{23}^2$ are the usual Mandelstam invariants. 
After rescaling $b$ by $|c|$, the $a$ and $c$ integrations both become elementary, and the $b$ integration can be done in the limit of $d\!\to\!4$.
With this, we find the all-plus helicity amplitude to be given by:
\vspace{-5pt}\eq{
\frac{\mi i\phantom{\mi}}{12{\cdot}16\pi^2}\frac{[12][34]}{\langle 12\rangle\langle34\rangle}
\left( \frac{s}{s+t}+\frac{s\,t}{(s+t)^2}\log\Big(\frac{s}{t}\Big)\right)+\text{cyclic}\,.\nonumber
\vspace{-5pt}}
Notice that logarithm nicely cancels in the sum, 
leaving the correct answer (see e.g.~\cite{Bern:1995db}),
\vspace{-5pt}\eq{\mathcal{A}^{1\text{-loop}}(\pl,\pl,\pl,\pl)=\mi\frac{1}{6}\frac{i}{16\pi^2}\frac{[12][34]}{\langle 12\rangle\langle34\rangle}\,.\vspace{-5pt}}

Regarding extensions to higher multiplicity and higher loops, it is worth mentioning that the $Q$-cut representation
can be combined with other modern techniques---for example, integral reduction and the use of integration-by-parts identities.

\vspace{-10pt}\section{Extensions to Higher Loops}\label{higher_loop_section}\vspace{-10pt}

To generalize the construction to two loops,
we begin by writing each Feynman diagram such that only
loop momenta $\ell_1,\ell_2$ or $(\ell_1\pl\ell_2)$ enter propagators.
We then separately partial-fraction-out the propagators of each three type.
More precisely, we introduce a three-parameter deformation $\ell_i\!\mapsto\!\ell_i\pl\eta_i$
where $\eta_1^2\!=\!z_1$, $\eta_2^2\!=\!z_2$ and $(\eta_1\pl\eta_2)^2\!=\!z_3$.
Partial-fractioning then expresses the amplitude in terms of its residues in $z_1,z_2,z_3$.

In each variable $z_i$ there are residues at both finite and infinite locations.
The residues with all three $z_i$ finite are immediately given by on-shell three-particle cuts, given by the following \mbox{$Q$-cut}:
\vspace{-17.5pt}\eq{\hspace{-17.5pt}\fwbox{65.pt}{\fig{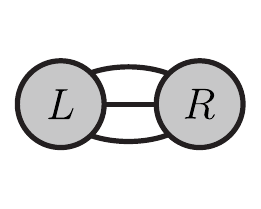}}\bbox{\Leftrightarrow}\frac{\mathcal{A}_L(\widetilde{\ell}_1,\widetilde{\ell}_2,
\mi\widetilde{\ell}_3)\mathcal{A}_R(\!\mi\widetilde{\ell}_1,\!\mi\widetilde{\ell}_2,\widetilde{\ell}_3)}{\ell_1^2\ell_2^2(\ell_1\pl\ell_2\pl P_L)^2},\vspace{-17.5pt}\label{two_loop_q_cuts_1}}
where $\tilde{\ell}_3\!\equiv\! \tilde{\ell}_1\pl\tilde{\ell}_2\pl P_L$ and the extra-dimensional components $\tilde{\ell}_i\!\equiv\!\ell_i\pl\eta_i$ 
are such that $\tilde{\ell}_{i}^2\!=\!0$ for $i\!=\!1,2,3$.
The residues at infinity require more work.
It is easy to see that residues with two or three  $z_i$ at infinity yield vanishing integrals and so can be discarded.
Residues with one at infinity, say $z_3$, represent degenerate topologies with disconnected loops.
For these terms we repeat the procedure introduced already at one-loop
and rescale $\ell_i\!\to\!\alpha_i\ell_i$ and expand by partial fractions separately in $\alpha_{1},\alpha_{2}$.
This yields a second type of \mbox{$Q$-cut}:
\vspace{-18pt}\begin{align}&\hspace{70pt}\fwbox{110pt}{\fig{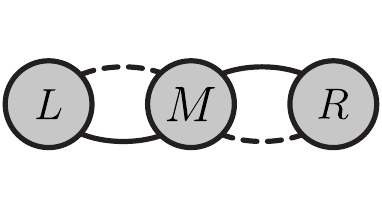}}\bbox{\Leftrightarrow}\label{two_loop_q_cuts_2}\\[-17.5pt]&\hspace{-3pt}\nonumber\frac{\mathcal{A}_L\hspace{-1pt}(\widetilde{\ell_1},\!\!\mi\widetilde{\ell}_1\!\mi\!P_L\hspace{-1pt})}{\ell_1^2(2\vecdot{\ell_1}{P_L}\!\pl\! P_L^2)}\hspace{-1pt}\mathcal{A}_M\hspace{-1pt}(\!\mi\widetilde{\ell}_1,\widetilde{\ell}_1\!\pl\!P_L,\!\!\mi\widetilde{\ell}_2,\widetilde{\ell}_2\!\pl\!P_R)\hspace{-1pt}\frac{\mathcal{A}_R\hspace{-1pt}(\widetilde{\ell}_2,\!\!\mi\widetilde{\ell}_2\!\mi\!P_R)}{\ell_2^2(2\vecdot{\ell_2}{P_R}\!\pl\!P_R^2)}.\\[-21.5pt]~\nonumber\vspace{-0pt}\end{align}
These tree amplitudes are evaluated using $\tilde{\ell_i}\!\equiv\!\alpha_i(\ell_i\pl\eta_i)$,
with $\eta_i^2\!=\!\mi\ell_i^2$ and $\alpha_{1,2}\!\equiv\!\mi P_{L,R}^2/(2\ell_{1,2}{\cdot}P_{L,R})$,
similar to the one-loop case. The amplitudes are projected onto the $0$th-order term in the Laurent expansion for large $\eta_1\!\cdot\!\eta_2$.

Intuitively the two \mbox{$Q$-cuts} above account for graphs with connected and disconnected loops, with
the extra-dimensional deformation and $\eta_1\!\cdot\!\eta_2\!\to\!\infty$ projection accomplishing a gauge-invariant separation between them.
According to our derivation, adding all \mbox{$Q$-cuts} (for all possible external leg insertions) will reproduce the correct integrated amplitude.
The contour for each \mbox{$Q$-cut} is determined by equations~(\ref{single_factor_pv})--(\ref{three_factor_pv}), which should be applied \emph{separately}
for each group of linear propagators---that is, those involving $(\vecdot{A}{\ell_1}\pl B)$, $(\vecdot{A}{\ell_2}\pl B)$ or $(\vecdot{A}{(\ell_1{+}\ell_2)}\pl B)$.

As a natural variation of the same technique, in a planar theory
one could restrict to a two-parameter deformation with $z_3\!=\!0$,
since no graph has more than one mixed propagator $1/(\ell_1\pl\ell_2\pl P)^2$.
The residues would then be related to the double-forward limits of trees, in theories where this limit can be defined~\cite{CaronHuot:2010zt}.

We believe that the \mbox{$Q$-cut}-construction generalizes straightforwardly to any loop order.
For example, using a $6$ parameter deformation with all $\eta_i\!\cdot\!\eta_j$ taken independent at three loops, we obtain the six \mbox{$Q$-cut} diagrams:
\vspace{-15pt}\eq{\hspace{-100pt}\raisebox{-47.05pt}{\includegraphics[scale=1]{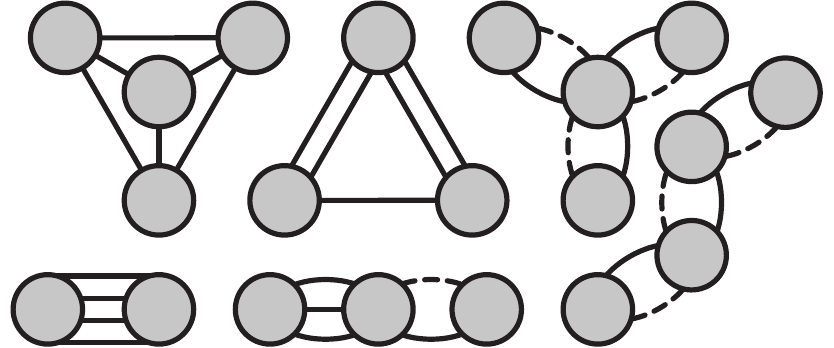}}\hspace{-100pt}\vspace{-10pt}\label{three_loop_q_cuts}}

\vspace{-10pt}\section{Summary and Conclusion}\label{conclusions_section}\vspace{-10pt}

The desire to represent loop amplitudes directly in terms of lower-loop amplitudes is motivated by the practical and conceptual advantages of eliminating explicit reference to the redundancies required by the Feynman-diagrammatic expansion. Recently, several such representations of loop amplitudes in terms of trees have appeared in the context of the scattering equation formalism. This is in part because this formalism makes it possible (at least for certain theories) to systematically regulate the divergences of tree amplitudes in the forward-limit~\cite{Baadsgaard:2015hia,He:2015yua}.\footnote{And in the case of planar $\phi^3$-theory, such a prescription exists to all orders~\cite{unpublishedCHY}, making it possible to write all $L$-loop amplitudes in terms of $L$-fold nested forward-limits of trees.} But it remains an important, open problem to systematically regulate the forward-limit divergences of amplitudes in general theories.

In this Letter we have described a new, ``$Q$-cut'' representation of loop amplitudes, derived from general field theory arguments and without any reference to forward limits. And this representation naturally extends to all orders of perturbation theory opening new possibilities for computation.

~\\[-10pt]\acknowledgments
We thank Zvi Bern, Lance Dixon, Peter Goddard and Lionel Mason for helpful discussions. This work has been supported in part by a MOBILEX grant from the Danish Council for Independent Research (JLB) and by Qiu-Shi funding and Chinese NSF funding under contracts No.11031005, No.11135006 and No.11125523 (BF). BF is grateful for the hospitality of the NBIA.

\end{document}